\documentstyle[prb,aps]{revtex}
\newcommand{\grad}{\nabla}
\newcommand{\be}{\begin{eqnarray}}
\newcommand{\ee}{\end{eqnarray}}
\newcommand{\n}{{\bf n}}
\newcommand{\D}{{\bf d}}
\newcommand{\V}{{\bf v}}
\begin{document}

\title{Velocity Distribution for Strings in Phase Ordering Kinetics}
\author{Gene F. Mazenko }
\address{The James Franck Institute and the Department of Physics \\
         The University of Chicago \\
         Chicago, Illinois 60637 }
\date{\today}
\maketitle
%
%
\begin{abstract}

The continuity equations expressing conservation
of  string defect charge can be used to find an explicit expression for the
string velocity field in terms of the order parameter in the case of an
$O(n)$ symmetric time-dependent Ginzburg-Landau model.
This expression for the velocity is used to find the
string velocity probability distribution in the case of phase-ordering
kinetics for a nonconserved
order parameter.  For long times $t$ after the quench, velocities
scale as $t^{-1/2}$.  There is a large velocity tail in the
distribution corresponding to annihilation of defects which goes
as $V^{-(2d+2-n)}$ for both point and string defects in $d$ spatial
dimensions.

\end{abstract}
\draft
\section{Introduction}

In recent work\cite{N.8} we discussed how one could use
conservation of topological
charge to study the statistics of velocities of point defects in
phase-ordering systems.  We were able to
identify the appropriate point-defect velocity field in terms of the order-
parameter field in the context of a $d$-dimensional $O(n)$
symmetric time-dependent Ginzburg-Landau (TDGL) model.
Using this expression
for the velocity field, for point particles ($n=d$), the probability
distribution for defect velocities was determined in
the case of the late-state phase ordering using the lowest-order approximation
in the perturbation expansion method developed in Ref.(\onlinecite{EXP}).
This analysis is extended here to the case of strings defects where
$n=d-1$.  The velocity probability distribution is worked out explicitly
for $n=d-1=1$ and $2$ and for the wall case $n=d-2=1$.  These results and
the results for all $n=d$ can
all be written in the form
\be
P({\bf V})=\Biggl(\frac{1}{\pi \bar{v}^{2}}\Biggr) ^{d/2}
\frac{\Gamma (\frac{d}{2})\Gamma (\frac{d}{2}+1)}
{\Gamma (\frac{n}{2})\Gamma (\frac{(d-n)}{2}+1)}
\left(\frac{\bar{v}}{V}\right)^{d-n}
\Biggl( 1+{\bf V}^{2}/\bar{v}^{2}\Biggr)^{-(d+2)/2}~~~.
\label{eq:1}
\ee
where the velocities scale with a factor
$\bar{v}\approx L(t)^{-1}$ where $L(t)\approx  t^{1/2}$ is the characteristic
scaling length
which grows with time t after the quench.
The result for $P({\bf V})$ indicates
that the probability of finding a defect with a large velocity decreases
with time.
There is a high-velocity
tail $V^{-(2d+2-n)}$ which corresponds to the annihilation of
defects and defect loops.
It is encouraging that this result agrees with  the large
$V$ tails obtained by Bray\cite{bvvt}  using rather general scaling ideas
for all $n\le d$.

The result obtained here for $P[{\bf V}]$ seems very simple.  Does it
correspond to experimental observation or the results of numerical
simulations?  Thus far there have been not direct tests.  It would seem
worthwhile to check the range of validity of the defect velocity
probability distribution given by Eq.(\ref{eq:1}).

\section{Problem Set Up}

We study an n-component non-conserved order-parameter field
$\psi_{\alpha} ( {\bf R} ,t)$ in $d$-spatial dimensions which
satisfies the TDGL equation
\be
\partial_{t} \psi_{\alpha} ( {\bf R} , t )=-\Gamma
{}~ \frac{\delta F [\vec { \psi} ]}
{\delta \psi_{\alpha} ( {\bf R} , t )}+\eta_{\alpha}( {\bf R} , t )
\label{eq:2}
\ee
where $F$ is an effective free-energy functional and
$\Gamma$ is a constant kinetic coefficient. We assume $F$ is of the
${\cal O}(n)$ symmetric square-gradient form
\be
F=\int~d ^ d R \left[ \frac{c}{2} ( \grad  \vec {\psi} ) ^
2
+V ( \vec {\psi} ) \right] ~~~,
\ee
where $ c > 0$ and $V ( \vec {\psi} )$
is chosen to be a degenerate double-well or wine-bottle potential.
This model is to be supplemented by random, uncorrelated,
initial conditions. We assume that there is a rapid temperature quench
from a high temperature to
zero temperature where
the noise $\eta_{\alpha}$ in (\ref{eq:2}) can be set to zero.
In the scalar case (n=1) such systems order through the growth of domains
separated by sharp walls. As time evolves these domains coarsen
and order grows to progressively longer length scales.
In the case of systems with continuous symmetry
($n > 1$) the disordering elements\cite{Mer,Kle} will depend on $n$ and
spatial dimensionality $d$.  Thus, for
example, for $n=d$ one has point defects (vortices or monopoles) while
for $n=d-1$ one has vortex lines or string-like objects.  For $n
> d $ there are no stable singular topological objects.

The main physics in phase-ordering  systems\cite{221,N.2,N.3} is
the interplay between
two characteristic lengths, a characteristic domain size $L(t)$, which
grows with time, and a defect dimension $\xi$ (interfacial width,
vortex core size, etc.). However,  at long enough times the single
length $L(t)$ dominates, $L(t)\gg\xi$, the morphological structure
looks self-similar under the rescaling of space and time and the order parameter
correlation function satisfies the scaling equation
\be
C_{\psi}({\bf R},t)\equiv\langle\vec {\psi}({\bf
R},t)\cdot \vec {\psi}(0,t)\rangle=\psi_{0}^{2}F(x)
\label{eq:3}
\ee
where $x\equiv R/L(t)$,  and $\psi _0$ is the magnitude
of $\vec{\psi}$ in the ordered state. The structure factor, the fourier
transform of $C_{\psi}({\bf R},t)$, satisfies
$\tilde{C}_{\psi}({\bf q},t)=L^{d}\psi_0^2 \tilde{F}(Q)$,
where $Q\equiv qL$ is a scaled wavenumber.
For pure systems
with short-range interactions
and a nonconserved order parameter
the growth law
is given by the Lifshitz-Cahn-Allen  result
$L ~\approx t ^ {1/2}$ for all $n$.   The two-dimensional XY model
\cite{RyKr,YPKH,ZGG,BRGL,LLK}
and the one-dimensional scalar model\cite{1dTDGL,1dTDGLa} appear
to be  interesting exceptions.
For large $Q$ and
$n\leq d$, due to defects, the structure factor obeys the generalized
Porod's Law,\cite{Porod0,DAB,Porod,OP,cs10}
$\tilde{F}(Q)\sim Q^{-(n+d) }$.  This reflects increasingly
weaker singularities in $F(x)$ for small $x$ as a function of $n$.
In the opposite limit, it appears that the large $x$ behavior can,
with proper definition of $x$, be put in the form
$F (x) \approx x^{-\nu}e^{-\frac{1}{2}x^{2}}$
where $\nu$ is a subdominant index\cite{EXP}.
The quantities discussed above are evaluated at equal times after
the quench.  In the two-time case one again has a scaling law
\cite{Furu89,FH,cs2} and the on-site correlation function  has the form
\be
< \vec{\psi} ( {\bf R} , \tau + t )\cdot\vec{ \psi} ( {\bf R} , t )~>
\approx~L(\tau) ^ {- \lambda } ~~~,
\label{eq:4}
\ee
for $\tau\gg t$, where $\lambda$ is
a nontrivial exponent which has been determined numerically and
theoretically\cite{EXP}
for a number of systems.

\section{Defect Dynamics for Point Particles}

Since a great deal is known about order-parameter correlations in
phase-ordering systems, attention has turned toward the
study\cite{MG,cs14} of the  statistics and
dynamics of the annihilating defects themselves.
The basic idea is that the positions of
defects are located by the zeros of the order parameter field
$\vec{ \psi}$, therefore the charged or signed density for
point defects
is given by
\be
\rho ({\bf R},t)=\delta(\vec{ \psi}({\bf R},t)){\cal D}({\bf R},t)
\label{eq:3.1}
\ee
where ${\cal D}$, associated with the change of variables
from the set of vortex positions to the field $\vec{\psi}$, is defined by:
\be
{\cal D}=\frac{1}{n!}\epsilon_{\mu_{1}\mu_{2}\ldots\mu_{n}}
\epsilon_{\nu_{1}\nu_{2}\ldots\nu_{n}}
\nabla_{\mu_{1}}\psi_{\nu_{1}}
\nabla_{\mu_{2}}\psi_{\nu_{2}}\ldots
\nabla_{\mu_{n}}\psi_{\nu_{n}}
\label{eq:det}
\ee
where $\epsilon_{\mu_{1}\mu_{2}\ldots\mu_{n}}$ is the
$n$-dimensional fully anti-symmetric tensor and
summation over repeated indices is implied here and below.
The unsigned  defect density, $n({\bf R},t)$,
is given by
\be
n ({\bf R},t)=\delta(\vec{ \psi}({\bf R},t))|{\cal D}({\bf R},t)| ~~~.
\label{eq:8}
\ee

It was shown in Ref.(\onlinecite{N.8}) that the
vortex charge density $\rho$ for point defects, defined by
Eq.(\ref{eq:3.1}),  satisfies a continuity equation of the form
\be
\dot{\rho} =-\nabla_{\mu_{1}}[\rho v_{\mu_{1}}]
\label{eq:9}
\ee
where the defect velocity field $v_{\mu_{1}}$  is given by
\be
{\cal D} v_{\mu_{1}}
=- \frac{1}{(n-1)!}\epsilon_{\mu_{1}\mu_{2}\ldots\mu_{n}}
\epsilon_{\nu_{1}\nu_{2}\ldots\nu_{n}}
\dot{\psi}_{\nu_{1}}
\nabla_{\mu_{2}}\psi_{\nu_{2}}\ldots
\nabla_{\mu_{n}}\psi_{\nu_{n}}
\label{eq:10}
\ee
where  ${\cal D}$ is defined by Eq.(\ref{eq:det}).  Thus one has an
explicit expression for the defect velocity field which can be expressed
strictly in terms of the order parameter and its spatial derivatives.
Remember that the TDGL equation of motion can be used to express
$\dot{\psi}_{\nu_{1}}$ in terms of $\psi$ and it spatial derivatives.

The expression given by Eq.(\ref{eq:10}) for the velocity is very
useful because it avoids the problem of having to specify the
positions of the defects explicitly.  The positions are
implicitly determined by the zeros of the order-parameter field.
The practical usefulness of Eq.(\ref{eq:10}) can be seen by
asking the question:  In the scaling regime of a phase-ordering
system with point defects, what is the probability of finding
a defect with a velocity ${\bf V}$?  This
probability distribution function is defined by
\be
\langle n\rangle P({\bf V})\equiv
\langle n \delta ({\bf V}-{\bf v}(\vec{\psi}))\rangle ~~~,
\ee
where ${\bf V}$ is a reference velocity and $n$ is the unsigned defect
density defined by Eq.(\ref{eq:8}). The calculated
$P({\bf V})$ is given by Eq.(\ref{eq:1}) with $n=d$.

Going further along these lines\cite{vvv}  we considered
the two
vortex velocity probability distribution,
$P[{\bf V}_{1},{\bf V}_{2},{\bf R}]$
which gives one the probability of finding
the velocity of one defect in the fixed presence of another defect a
known distance away with a known velocity.
Clearly $P[{\bf V}_{1},{\bf V}_{2},{\bf R}]$
contains a
tremendous amount of information about the dynamics of point defects.
The  physical results from the calculation of this quantity,
carried out in detail
for $n=d=2$ in REf.(\onlinecite{vvv}, are relatively simple to state.
The probability distribution is a function only of the scaled
velocities $\vec{u}_{i}=\vec{V}_{i}/\bar{v}$ for $i=1$ or $2$, and
the scaled separation $\vec{x}=\vec{R}/L(t)$.
The characteristic velocity
$\bar{v}$ is the same quantity that appears in $P({\bf V})$.
For a given scaled separation  $x$ ,
the most probable configuration
corresponds, as expected, to a state with zero total velocity and
a nonzero
relative velocity only along the axis connecting the vortices:
${\bf V}_{1}=-{\bf V}_{2}\equiv v\hat{x}$. Moreover there is
a definite most probable nonzero value for $v=v_{max}$ for a given
value of $x$. The most
striking feature of these results is that for small $x$ the
most probable velocity goes as $v_{max}=\kappa /R$
where $R$ is the unscaled separation between the vortices and
$\kappa =2.19$ in dimensionless units defined in Ref.(\onlinecite{vvv}).
The result giving $v_{max}$ inversely proportional to $R$ is consistent with
overdamped dynamics where the relative velocity of the two vortices
is proportional to the
force which in turn is the derivative of a potential which is logarithmic
in the separation distance.
Since there is low probability\cite{MG,vaok} of finding like-signed
vortices at short
distances, our results giving the velocity as a function
of separation distance should be interpreted in terms of annihilating
vortex-antivortex pairs and is in agreement with the general scaling
ideas proposed by Bray\cite{bvvt}.

\section{Continuity Equations}

Let us investigate the existence of a local statement of topological
charge in the general case of $n\le d$.  The first step is to
introduce the appropriate density for topological charge.  In the case
of point particles
the conserved density is the charge density given by
Eq.(\ref{eq:3.1}). The next obvious extension\cite{ctg}
is to string defects where $n=d-1$ and the
the defect line density is given
by
\be
\rho _{s_{1}}
= \delta (\vec{ \psi}){\cal D}_{s_{1}}
\label{eq:20d}
\ee
where
\be
{\cal D} _{s_{1}}
=\frac{1}{n!}\epsilon_{s_{1} \mu_{1} \mu_{2} ...\mu_{n}}
\epsilon_{\nu_{1} \nu_{2} \ldots\nu_{n}}
\nabla_{\mu_1}\psi_{\nu_{1}}
\nabla_{\mu_2}\psi_{\nu_{2}}....\nabla_{\mu_n}\psi_{\nu_{n}} ~~~.
\label{eq:20}
\ee
and the $s$ and $\mu$ range from $1$ to $d$ and the $\nu$ range
from $1$ to $n$ and there is summation over repeated indices.
The generalization of Eqs.(\ref{eq:9}) and (\ref{eq:20d}) to all
$n\le d$ is given by
\be
\rho _{s_{1}s_{2}\ldots s_{d-n}}
= \delta (\vec{ \psi}){\cal D}_{s_{1}s_{2}\ldots s_{d-n}}
\label{eq:20a}
\ee
where
\be
{\cal D} _{s_{1}s_{2}\ldots s_{d-n}}
=\frac{1}{n!}
\epsilon_{s_{1}s_{2}\ldots s_{d-n} \mu_{1} \mu_{2}\ldots\mu_{n}}
\epsilon_{\nu_{1} \nu_{2} \ldots\nu_{n}}
\nabla_{\mu_1}\psi_{\nu_{1}}
\nabla_{\mu_2}\psi_{\nu_{2}}....\nabla_{\mu_n}\psi_{\nu_{n}} ~~~.
\label{eq:15}
\ee
It is then straightforward
to show that  ${\cal D} _{s_{1}s_{2}\ldots s_{d-n}}$ itself satisfies
a continuity equation given by
\be
\dot{{\cal D}}_{s_{1}s_{2}\ldots s_{d-n}}
=\nabla_{\mu_{1}}J_{ s_{1}s_{2}\ldots s_{d-n}\mu_{1}} ~~~,
\label{eq:16}
\ee
with the current defined by
\be
J_{s_{1}s_{2}\ldots s_{d-n}\mu_{1}}
=\epsilon_{s_{1}s_{2}\ldots s_{d-n}\mu_{1}\mu_{2} \dots \mu_{n}}
g_{\mu_{2}\mu_{3} \ldots \mu_{n}} ~~~,
\ee
and
\be
g_{\mu_{2}\mu_{3} \ldots \mu_{n}}
=\frac{1}{(n-1)!}
\epsilon_{\nu_{1}\nu_{2} \dots \nu_{n}}
\dot{\psi}_{\nu_{1}}
\nabla_{\mu_{2}}\psi_{\nu_{2}} \dots
\nabla_{\mu_{n}}\psi_{\nu_{n}}.
\ee
To obtain the continuity equation
for $\rho _{s_{1}s_{2}\ldots s_{d-n}}$, we need a second identity.
Consider the quantity
\be
J_{s_{1}s_{2}\ldots s_{d-n}\mu_{1}}
\nabla_{\mu_{1}}\psi_{\nu} =
\epsilon_{s_{1}s_{2}\ldots s_{d-n}
\mu_{1}\mu_{2} \dots \mu_{n}}
g_{\mu_{2}\mu_{3} \ldots \mu_{n}}\nabla_{\mu_{1}}\psi_{\nu}
\nonumber
\ee
\be
=\epsilon_{s_{1}s_{2}\ldots s_{d-n}
\mu_{1}\mu_{2} \dots \mu_{n}}\epsilon_{\nu_{1}\nu_{2} \dots \nu_{n}}
\nabla_{\mu_{1}}\psi_{\nu}\nabla_{\mu_{2}}\psi_{\nu_{2}} \dots
\nabla_{\mu_{n}}\psi_{\nu_{n}} ~~~.
\label{eq:21}
\ee
The key observation is that the right-hand-side of Eq.(\ref{eq:21}) has
a factor which can be written in the form:
\be
\epsilon_{s_{1}s_{2}\ldots s_{d-n}
\mu_{1}\mu_{2} \dots \mu_{n}}\nabla_{\mu_{1}}
\psi_{\nu}\nabla_{\mu_{2}}\psi_{\nu_{2}} \dots
\nabla_{\mu_{n}}\psi_{\nu_{n}}
=\epsilon_{\nu\nu_{2} \dots \nu_{n}} Q_{s_{1}s_{2}\ldots s_{d-n}}
\label{eq:22}
\ee
where $Q_{s_{1}s_{2}\ldots s_{d-n}}$ is
determined by multiplying Eq.(\ref{eq:22}) by
$\epsilon_{\nu\nu_{2} \dots \nu_{n}}$
and summing over all the $\nu$'s.  We easily obtain, remembering
Eq.(\ref{eq:15}),
\be
n!~Q_{s_{1}s_{2}\ldots s_{d-n}}=n! ~ {\cal D}_{s_{1}s_{2}\ldots s_{d-n}} ~~~.
\ee
Putting this result back into Eq.(\ref{eq:21}) gives
\be
J_{s_{1}s_{2}\ldots s_{d-n}\mu_{1}}
\nabla_{\mu_{1}}\psi_{\nu} =
\frac{1}{(n-1)!}\dot{\psi}_{\nu_{1}}\epsilon_{\nu_{1}\nu_{2} \dots \nu_{n}}
\epsilon_{\nu\nu_{2} \dots \nu_{n}} {\cal D}_{s_{1}s_{2}\ldots s_{d-n}}
\nonumber
\ee
\be
=\frac{1}{(n-1)!}\dot{\psi}_{\nu_{1}}\delta_{\nu_{1},\nu}(n-1)!
{\cal D}_{s_{1}s_{2}\ldots s_{d-n}}
\nonumber
\ee
\be
=\dot{\psi}_{\nu}{\cal D}_{s_{1}s_{2}\ldots s_{d-n}} ~~~.
\label{eq:24}
\ee
Taking the time derivative of $\rho_{s_{1}s_{2}\ldots s_{d-n}}$
gives, using Eqs. (\ref{eq:10}) and (\ref{eq:24}),
\be
\dot{\rho}_{s_{1}s_{2}\ldots s_{d-n}}
=\frac{\partial \delta (\vec{\psi})}{\partial \psi_{\nu}}
\dot{\psi}_{\nu}~{\cal D}_{s_{1}s_{2}\ldots s_{d-n}}
+\delta (\vec{\psi})\dot{{\cal D}}_{s_{1}s_{2}\ldots s_{d-n}}
\nonumber
\ee
\be
=\frac{\partial \delta (\vec{\psi})}{\partial \psi_{\nu}}
J_{s_{1}s_{2}\ldots s_{d-n}\mu_{1}}
\nabla_{\mu_{1}}\psi_{\nu}+\delta (\vec{\psi})\nabla_{\mu_{1}}
J_{s_{1}s_{2}\ldots s_{d-n}\mu_{1}}
\nonumber
\ee
and finally we obtain the desired continuity equation
\be
\dot{\rho}_{s_{1}s_{2}\ldots s_{d-n}}
=\nabla_{\mu_{1}}\left(\delta
(\vec{\psi})J_{s_{1}s_{2}\ldots s_{d-n}\mu_{1}}\right)
{}.
\label{eq:19}
\ee

For the simplest case of point defects ($n=d$), Eq.(\ref{eq:19})
can be put into the conventional form given by Eq.(\ref{eq:9}) with
\be
J_{\mu_{1}}=
- v_{\mu_{1}}{\cal D}~~~,
\ee
and the velocity given by Eq.(\ref{eq:10}).

Let us turn next to the case of strings where the line density is a
vector $\rho_{\mu_{1}}$ and the current is a two-component tensor,
\be
J_{s_{1}\mu_{1}}=
\epsilon_{s_{1}
\mu_{1}\mu_{2} \dots \mu_{n}}
g_{\mu_{2}\mu_{3} \ldots \mu_{n}} ~~~.
\ee
Clearly $J_{s_{1}\mu_{1}}$ is anti-symmetric in its subscripts.  Since
we expect the instantaneous velocity to be orthogonal to the local
orientation of the string, we can define the velocity via
\be
J_{\alpha\beta} =v_{\alpha}{\cal D}_{\beta}-v_{\beta}{\cal D}_{\alpha}
{}~~~.
\ee
Dotting the vector $\vec{{\cal D}}$ into this expression gives the
result
\be
v_{\alpha}=\frac{1}{\vec{{\cal D}}^{2}}J_{\alpha\beta}{\cal D}_{\beta}
{}~~~,
\ee
where we have taken advantage of the fact that $\vec{v}$ and $\vec{{\cal D}}$
are orthogonal:
\be
\vec{{\cal D}}\cdot \vec{v}
={\cal D}_{\alpha} \frac{1}{\vec{{\cal D}}^{2}}J_{\alpha\beta}{\cal D}_{\beta}
=0 ~~~.
\ee
The velocity field for strings can be written in the form
\be
v_{s_{1}}=\frac{1}{\vec{{\cal D}}^{2}}{\cal D}_{\mu_{1}}
\epsilon_{s_{1}
\mu_{1}\mu_{2} \dots \mu_{n}}
g_{\mu_{2}\mu_{3} \ldots \mu_{n}}
\label{eq:29}
\ee
for general $n$.

Let us check this result and its sign for the simplest case of
$n=1$, $d=2$.  The vector $\vec{{\cal D}}$ in this case takes the simple
form
\be
{\cal D}_{\mu_{1}}=\epsilon_{\mu_{1}\mu_{2}}\nabla_{\mu_{2}}\psi ~~~,
\ee
\be
g=\dot{\psi} ~~~,
\ee
and
\be
v_{s_{1}}=\frac{1}{(\nabla\psi )^{2}}\epsilon_{\mu_{1}\mu_{2}}
\nabla_{\mu_{2}}\psi\epsilon_{s_{1}\mu_{1}} \dot{\psi}
=-\frac{\nabla_{\mu_{1}}\psi}{(\nabla\psi )^{2}}\dot{\psi} ~~~.
\ee
Consider a circular loop of string of radius $R(t)$ where the order
parameter near the interface formed by the loop is given in polar
coordinates in the
form
\be
\psi (\vec{r}) =A (r-R(t)) ~~~.
\label{eq:30}
\ee
where $A$ is an overall constant amplitude.
We then need the derivatives
\be
\nabla_{\mu_{1}}\psi=A ~\hat{r}_{\mu_{1}}
\ee
and
\be
\dot{\psi}=-\dot{R}(t)A
\ee
to obtain the velocity
\be
v_{s_{1}}=\frac{1}{A ^{2}}\dot{R}(t)\hat{r}_{s_{1}}A^{2}
=\dot{R}(t)\hat{r}_{s_{1}}
\ee
as expected.  Typically we will substitute for $\dot{\psi}$ using
the equation of motion and use the defect locating $\delta$-function
to set
\be
\dot{\psi}=\Gamma c \nabla^{2}\psi
\label{eq:37}
\ee
and
\be
v_{s_{1}}
=-\frac{(\nabla_{s_{1}}\psi)}{(\nabla\psi )^{2}}
\Gamma c \nabla^{2}\psi ~~~.
\ee
Using the same ansatz given by Eq.(\ref{eq:30}) and remembering that
the expression for the velocity is multiplied by a $\delta$-function
setting $r=R(t)$, leads to the result:
\be
\nabla^{2}\psi =\frac{A}{r}=\frac{A}{R(t)}
{}~~~,
\ee
and we obtain the Lifshitz-Cahn-Allen\cite{221,N.2,N.3} result,
\be
v_{s_{1}} =-\Gamma c \frac{1}{R(t)} \hat{r}_{s_{1}}~~~,
\ee
which tells us that the circular domain is shrinking and
$R(t)\approx t^{1/2}$.

For $n=2$ and $d=3$ we have explicitly the results reported in
Ref.(\onlinecite{ctg})
\be
\vec{{\cal D}}=\frac{1}{2}\epsilon_{\nu_{1}\nu_{2}}
\left(\vec{\nabla} \psi_{\nu_{1}}\times \vec{\nabla} \psi_{\nu_{2}}\right)
\ee
\be
\vec{g}=\epsilon_{\nu_{1}\nu_{2}}\dot{\psi}_{\nu_{1}}\vec{\nabla} \psi_{\nu_{2}}
\ee
and
\be
\vec{v}=\frac{1}{\vec{{\cal D}}^{2}} \vec{{\cal D}}\times\vec{g} ~~~.
\ee

\section{Brief Discussion of Walls}

Let us briefly discuss how things develop as one
attempts to go further and increase $d-n$ to 2 and the case of walls.
We must in this case deal with the quantities
\be
{\cal D} _{s_{1}s_{2}}
=\frac{1}{n!}\epsilon_{s_{1}s_{2} \mu_{1} \mu_{2} ...\mu_{n}}
\epsilon_{\nu_{1} \nu_{2} \ldots\nu_{n}}
\nabla_{\mu_1}\psi_{\nu_{1}}
\nabla_{\mu_2}\psi_{\nu_{2}}....\nabla_{\mu_n}\psi_{\nu_{n}}
\label{eq:20c}
\ee
and
\be
J_{s_{1}s_{2}\mu_{1}}
=\epsilon_{s_{1}s_{2}
\mu_{1}\mu_{2} \dots \mu_{n}}
g_{\mu_{2}\mu_{3} \ldots \mu_{n}}~~~.
\ee
If we recall the definitions of the velocity field for point defects
\be
J_{\mu_{1}}=-  v_{\mu_{1}}{\cal D}
\ee
and for string defects
\be
J_{\alpha\beta} =v_{\alpha}{\cal D}_{\beta}-v_{\beta}{\cal D}_{\alpha}~~~,
\ee
then it is natural to write for walls
\be
J_{s_{1}s_{2}\mu_{1}}=-{\cal D}_{s_{1}s_{2}}v_{\mu_{1}}
-{\cal D}_{\mu_{1}s_{1}}v_{s_{2}}-{\cal D}_{s_{2}\mu_{1}}v_{s_{1}} ~~~.
\label{eq:48}
\ee
This expression builds in the antisymmetry of $J_{s_{1}s_{2}\mu_{1}}$.

If one restricts the discussion to the physically most relevant case of
$n=1,~d=3$, then one can use the result
\be
{\cal D}_{s_{1}s_{2}}=\epsilon_{s_{1}s_{2}\mu_{3}}
\nabla_{\mu_{3}} \psi
\ee
to show that the velocity field is given by
\be
v_{\mu_{1}}=-\frac{1}{{\cal D}^{2}}{\cal D}_{s_{1}s_{2}}
\epsilon_{s_{1}s_{2}\mu_{3}}
\nabla_{\mu_{1}}\dot{\psi}
\ee
where
\be
{\cal D}^{2}=\left(\vec{\nabla}\psi \right)^{2} ~~~.
\ee
Further straight-forward manipulation leads to the final result
\be
v_{\mu}=-\frac{\nabla_{\mu}\psi }{\left(\vec{\nabla}\psi \right)^{2}}
\dot{\psi} ~~~.
\label{eq:48a}
\ee
The key result here is that $\vec{v}$ is orthogonal to
${\cal D}_{s_{1},s_{2}}$:
\be
v_{\mu_{1}}{\cal D}_{\mu_{1},s_{3}}=
\frac{\dot{\psi}}{\left(\vec{\nabla}\psi \right)^{2}}
\nabla_{\mu_{1}}\psi ~\epsilon_{\mu_{1},s_{3},\mu_{3}}\nabla_{\mu_{3}}\psi =0~~~.
\ee
It remains to be seen if Eq.(\ref{eq:48}) serves as a useful definition of
the velocity field for walls with $n>1$.
Notice that the results for a scalar order parameter $(n=1)$ can all
be written in the form of Eq.(\ref{eq:48a}) for $d=1,2$ and $3$.

\section{Evaluation of $P[{\bf V}]$ for Strings}

Our interest here is in determining the defect velocity probability
distribution, $P[{\bf V}]$, for strings.
Again we  use the auxiliary field method\cite{OJK,YJ,TUG,cs1,YOS} which
has been successful in
determining the scaling function for the order-parameter correlation
function in a perturbation theory expansion.  We evaluate $P$ here to
lowest order in this expansion where the auxiliary field can be treated as
a gaussian field.  The first step in this theory is to express
the order parameter in terms of
an auxiliary field $\vec{m}$.
For our purposes here the important result is that near a
charge one vortex core the order parameter is linear in the auxiliary
field
\be
\vec{\psi}(\vec{m})=A\vec{m}+{\cal O}(m^{3})~~~.
\ee
It is then
easy to show that one can replace $\vec{\psi}$ by $\vec{m}$ in the
expression for $\vec{v}$  given by Eq.(\ref{eq:29})  and in the
expression for the string-charge density
\be
\rho_{\alpha}(\vec{\psi})=\rho_{\alpha}(\vec{m}) ~~~.
\ee

We then want to determine the string-velocity probability distribution
\be
\langle |\vec{\rho}|\rangle P({\bf V})\equiv
\langle |\vec{\rho}(\vec{\psi})| \delta ({\bf V}-{\bf v}(\vec{\psi}))\rangle ~~~,
\nonumber
\ee
\be
=\langle |\vec{\rho}(\vec{m})| \delta ({\bf V}-{\bf
v}(\vec{m}))\rangle ~~~.
\ee
One can determine $P({\bf V})$ by first
evaluating the more general probability distribution
\be
G(\xi ,\vec{b})=\langle \delta (\vec{m} )
\delta (\xi_{\mu}^{\nu}-\nabla_{\mu}m_{\nu })
\delta (\vec{b}-\nabla^{2}\vec{m})\rangle
\ee
since
\be
n_{0}P({\bf V})=
\int d^{n}b \prod_{\mu=1}^{d}\prod_{\nu=1}^{n}d\xi_{\mu}^{\nu}
|\vec{{\cal D}} (\xi )|
\delta (\vec{V}-\vec{v}(\vec{b},\xi ))G(\xi ,\vec{b})
\label{eq:57}
\ee
where
\be
v_{\mu}(\vec{b},\xi ))=\frac{\Gamma c}{\vec{{\cal D}}^{2}}
\frac{1}{(n-1)!}\epsilon_{\mu_{1}s_{1}\mu_{2}\ldots\mu_{n}}
{\cal D}_{s_{1}}(\xi )
\epsilon_{\nu_{1}\nu_{2}\ldots\nu_{n}}
b_{\nu_{1}}
\xi_{\mu_{2}}^{\nu_{2}}\ldots
\xi_{\mu_{n}}^{\nu_{n}}
\ee
with
\be
{\cal D}_{s_{1}}(\xi )=\frac{1}{n!}\epsilon_{s_{1}\mu_{1}\mu_{2}\ldots\mu_{n}}
\epsilon_{\nu_{1}\nu_{2}\ldots\nu_{n}}
\xi_{\mu_{1}}^{\nu_{1}}
\xi_{\mu_{2}}^{\nu_{2}}\ldots
\xi_{\mu_{n}}^{\nu_{n}}
\label{eq:60}
\ee
and
\be
n_{0}=\langle |\vec{\rho}(\vec{m}|\rangle~~~.
\ee
We have assumed that the quench is to
zero temperature so
that the noise can be set to zero and we can use Eq.(\ref{eq:37}).
The gaussian average determining $G(\xi ,\vec{b})$ is relatively
straightforward to evaluate\cite{30} and is given by
\be
G(\xi ,\vec{b})=\frac{1}{(2\pi S_{0})^{n/2}}
\frac{e^{-\frac{1}{2\bar{S}_{4}}\vec{b}^{2}}}{(2\pi \bar{S}_{4})^{n/2}}
\frac{1}{(2\pi \bar{S}_{2})^{nd/2}}
exp\biggl[-\frac{1}{2\bar{S}_{2}}\sum_{\mu
,\nu}(\xi_{\mu}^{\nu})^{2}\biggr]
\label{eq:62}
\ee
where
\be
S_{0}=\frac{1}{n}\langle \vec{m}^{2}\rangle\approx L^{2}
\ee
\be
\bar{S}_{2}=\frac{1}{dn}\langle (\nabla \vec{m})^{2}\rangle
\approx L^{0}
\ee
and
\be
\bar{S}_{4}=\frac{1}{n}\langle (\nabla^{2}\vec{m})^{2}\rangle
-\frac{\biggl(d\bar{S}_{2}\biggr)^{2}}{S_{0}}
\approx L^{-2} ~~~.
\ee
The quantities $S_{0}, \bar{S}_{2},\bar{S}_{4}$ are determined from the
theory
for the order parameter correlation function and discussed further below.

The problem then reduces to evaluating the $\vec{b}$ and $\xi $
integrations in the integral given by Eq.(\ref{eq:57}) using the result
for $G(\xi ,\vec{b})$ given by Eq.(\ref{eq:62}).  We proceed by first
doing the integration over $\vec{b}$.  This is facilitated by first
defining the matrix $M_{\mu}^{\nu}$ via
\be
v_{\mu}=\Gamma c M_{\mu}^{\nu}b_{\nu}
\ee
and
\be
M_{\mu}^{\nu}=\frac{1}{(n-1)!}\frac{1}{\vec{{\cal D}}^{2}(\xi )}
\epsilon_{\mu s_{1}\mu_{2}\ldots\mu_{n}}{\cal D}_{s_{1}}(\xi )
\epsilon_{\nu\nu_{2}\ldots\nu_{n}}
\xi_{\mu_{2}}^{\nu_{2}}\ldots\xi_{\mu_{n}}^{\nu_{n}}
{}~~~~.
\label{eq:67}
\ee
Clearly the quantity $\vec{{\cal D}}^{2}$ is important in the development
and is discussed in Appendix A.  The matrix $M_{\mu}^{\nu}$ is discussed
in Appendix B.
Then we use the integral representation for the $\delta$-function
and find that we can evaluate the integral over $\vec{b}$ in terms
of standard displaced gaussian integrals with the results:
\be
n_{0}P[{\bf V}]=\frac{1}{\left(2\pi S_{0}\right)^{n/2}}
\int  \prod_{\mu=1}^{d}\prod_{\nu=1}^{n}d\xi_{\mu}^{\nu}
|\vec{{\cal D}}(\xi )|
\frac{1}{(2\pi \bar{S}_{2})^{nd/2}}
exp\biggl[-\frac{1}{2\bar{S}^{(2)}}\sum_{\mu
,\nu}(\xi_{\mu}^{\nu})^{2}\biggr] J(\xi )
\label{eq:68}
\ee
where
\be
J(\xi )=\int ~\frac{d^{d}k}{(2\pi)^{d}}e^{i\vec{k}\cdot\vec{V}}
J_{k}(\xi )
\ee
and
\be
J_{k}(\xi )= \int ~\frac{d^{d}b}{(2\pi\bar{S}_{4})^{n/2}}
e^{-i\vec{k}\cdot\vec{v}(\vec{b},\xi )} e^{-\frac{1}{2\bar{S}_{4}}
\vec{b}^{2}}
\nonumber
\ee
\be
= \int ~\frac{d^{n}b}{(2\pi\bar{S}_{4})^{n/2}}
e^{-ik_{\mu}\Gamma c M_{\mu}^{\nu}b_{\nu}} e^{-\frac{1}{2\bar{S}_{4}}
\vec{b}^{2}}
\nonumber
\ee
\be
=exp\left[-\frac{1}{2}\bar{S}_{4}(\Gamma c)^{2}
k_{\alpha}k_{\beta}M_{\alpha}^{\nu}M_{\beta}^{\nu}\right] ~~~.
\ee
Sums over $\alpha$ and $\beta$ range from $1$ to $d$, those over
$\nu$'s from $1$ to $n$.
We then have the apparently Gaussian integral over $\vec{k}$:
\be
J(\xi )=\int ~\frac{d^{d}k}{(2\pi)^{d}}e^{i\vec{k}\cdot\vec{V}}
exp\left[-\frac{1}{2}\bar{S}_{4}(\Gamma c)^{2}
k_{\alpha}k_{\beta}\bar{M}_{\alpha\beta}\right]
\label{eq:71}
\ee
where
\be
\bar{M}_{\alpha\beta} =M_{\alpha}^{\nu}M_{\beta}^{\nu}
\ee
is a symmetric $d\times d$ matrix.
Since we have the property
\be
{\cal D}_{\alpha}M_{\alpha}^{\nu}=0 ~~~,
\ee
we see that the matrix $\bar{M}_{\alpha\beta}$ has a zero eigenvalue
corresponding to an eigenfunction in the $\vec{{\cal D}}$ direction:
\be
{\cal D}_{\alpha}\bar{M}_{\alpha\beta}=0 ~~~.
\ee
In order to carry out the integral in Eq.(\ref{eq:71}),
we must set up a coordinate
system which singles out the $\vec{{\cal D}}$ direction and $n$ orthogonal
directions.  Thus we construct an orthonormal set
$(\hat{{\cal D}}, \hat{\xi}^{(s)})$, for $s=1,2\ldots,n$, which satisfy
\be
\hat{\xi}^{(s)}_{\alpha}\hat{{\cal D}}_{\alpha}=0
\ee
\be
\hat{\xi}^{(s)}_{\alpha}\hat{\xi}^{(s')}_{\alpha}=\delta_{ss'}~~~.
\ee
In Appendix C we show that we can write
\be
\hat{\xi}^{(s)}_{\alpha}=\sum_{\nu}A_{s\nu}\xi^{\nu}_{\mu}
\ee
and it can be shown generally, see Appendix C, that
\be
(det ~ A)(det ~\tilde{A})=\frac{1}{det~N}=\frac{1}{\vec{{\cal D}}^{2}}
\ee
where $N$ is the $n \times n$ matrix
\be
N_{\nu\nu'}=\xi_{\mu}^{\nu}\xi_{\mu}^{\nu'} ~~~.
\ee
We then make the change of variables in the integral
Eq.(\ref{eq:71}) from
$\vec{k}$ to
\be
k_{\alpha}=\sum_{\nu 1}^{n}t_{\nu}\hat{\xi}^{(\nu )}_{\alpha}
+t_{d}\hat{{\cal D}}_{\alpha}
\ee
which clearly has a Jacobian of one.  The integral of interest is then
given by
\be
J(\xi )=\int ~\frac{d^{d}t}{(2\pi)^{d}}
e^{it_{d}\hat{{\cal D}}\cdot\vec{V}}
e^{it_{\nu}\hat{\xi}_{\alpha}^{(\nu )}V_{\alpha}}
exp\left[-\frac{\bar{S}_{4}}{2}(\Gamma c)^{2}
t_{\nu}t_{\nu'}Q_{\nu\nu'}\right]
\ee
where $\nu$ and $\nu'$ range from $1$ to $n$.  One can then do the
$t_{d}$ integration to obtain a $\delta$-function.  The rest of the
integrations are over gaussian fields governed by the matrix
\be
Q_{\nu\nu'}=\hat{\xi}_{\alpha}^{(\nu )}\hat{\xi}_{\beta}^{(\nu' )}
\bar{M}_{\alpha\beta}
\ee
which does not possess any zero eigenvalues.  Evaluating the standard
gaussian integral leads to  the result:
\be
J(\xi )=\delta (\hat{{\cal D}}\cdot\vec{V})
\frac{1}{(2\pi\bar{S}_{4}(\Gamma c)^{2})^{n/2}}\frac{1}{(det ~ Q)^{1/2}}
exp\left[-\frac{1}{2\bar{S}_{4}(\Gamma c)^{2}}V_{\nu}V_{\nu'}
\left(Q^{-1}\right)_{\nu\nu'}\right]
\label{eq:83}
\ee
where the $n$-vector $V_{\nu}$ is defined by
\be
V_{\nu}=\hat{\xi}_{\alpha}^{(\nu )}V_{\alpha}~~~.
\label{eq:84}
\ee
Note, if we insert Eq.(\ref{eq:84}) into Eq.(\ref{eq:83}), we
see that we need only the matrix
\be
R_{\alpha\beta}=\hat{\xi}_{\alpha}^{(\nu )}\hat{\xi}_{\beta}^{(\nu' )}
\left(Q^{-1}\right)_{\nu\nu'}
\label{eq:85}
\ee
and
\be
J(\xi )=\frac{\delta (\hat{{\cal D}}\cdot\vec{V})}{(\Gamma c)^{n}}
\frac{1}{(2\pi\bar{S}_{4})^{n/2}}\frac{1}{(det ~ Q)^{1/2}}
exp\left[-\frac{1}{2\bar{S}_{4}(\Gamma c)^{2}}V_{\alpha}V_{\beta}
R_{\alpha\beta}\right]~~~.
\ee
We show in Appendix D that the matrix $R$, defined by Eq.(\ref{eq:85}),
can be put into the very simple form
\be
R_{\alpha\beta}=\xi_{\alpha}^{\nu}\xi_{\beta}^{\nu} ~~~,
\ee
and finally $J(\xi )$ is given by:
\be
J(\xi )=\frac{\delta (\hat{{\cal D}}\cdot\vec{V})}{(\Gamma c)^{n}}
\frac{1}{(2\pi\bar{S}_{4})^{n/2}}|\vec{{\cal D}}|
exp\left[-\frac{1}{2\bar{S}_{4}(\Gamma c)^{2}}V_{\alpha}V_{\beta}
\xi_{\alpha}^{\nu}\xi_{\beta}^{\nu}\right] ~~~.
\ee
Inserting this result into Eq.(\ref{eq:68}) gives
\be
n_{0}P[{\bf V}]=\frac{1}{\left(2\pi S_{0}2\pi \bar{S}_{4}\right)^{n/2}}
\int  \prod_{\mu=1}^{d}\prod_{\nu=1}^{n}d\xi_{\mu}^{\nu}
\vec{{\cal D}}^{2}(\xi )
\frac{\delta (\hat{{\cal D}}\cdot\vec{V})}{(\Gamma c)^{n}}
\frac{1}{(2\pi \bar{S}_{2})^{nd/2}}
e^{-\frac{1}{2}A(\xi )}
\ee
where
\be
A(\xi )=
\frac{1}{\bar{S}_{2}}\sum_{\nu =1}^{n}\sum_{\alpha =1}^{d}
(\xi_{\alpha}^{\nu})^{2}
+\frac{1}{\bar{S}_{4}(\Gamma c)^{2}}\sum_{\nu =1}^{n}\sum_{\alpha ,\beta =1}^{d}
V_{\alpha}V_{\beta}
\xi_{\alpha}^{\nu}\xi_{\beta}^{\nu} ~~~.
\ee
If we  make the rescalings
\be
\xi_{\mu}^{\nu}\rightarrow \sqrt{\bar{S}_{2}}\xi_{\mu}^{\nu}
\label{eq:91}
\ee
and
\be
V_{\alpha} \rightarrow \Gamma c\sqrt{\frac{\bar{S}_{4}}{\bar{S}_{2}}}
\tilde{V}_{\alpha} ~~~,
\label{eq:92}
\ee
then
\be
\vec{{\cal D}}\rightarrow \left(\bar{S}_{2}\right)^{n/2}\vec{{\cal D}}
\ee
and we obtain
\be
n_{0}P[{\bf V}]=
\frac{1}{\left(2\pi \right)^{n}}\frac{1}{(\Gamma c)^{d}}
\sqrt{\frac{\bar{S}_{2}}{\bar{S}_{4}}}
\left(\frac{\bar{S}_{2}^{2}}{S_{0} \bar{S}_{4}}\right)^{n/2}
I(\tilde{V})
\label{eq:94}
\ee
where the dimensionless integral $I(\tilde{V})$ is defined by:
\be
I(\tilde{V})=
\int  \prod_{\mu=1}^{d}\prod_{\nu=1}^{n}d\xi_{\mu}^{\nu}~
\vec{{\cal D}}^{2}(\xi )\delta (\hat{{\cal D}}(\xi )\cdot\vec{\tilde{V}})
\frac{e^{-\frac{1}{2}A_{0}(\xi )}}{(2\pi )^{nd/2}}
\ee
with
\be
A_{0}(\xi )=
\sum_{\nu =1}^{n}\left[\sum_{\alpha =1}^{d}
(\xi_{\alpha}^{\nu})^{2}
+\sum_{\alpha ,\beta =1}^{d}
\tilde{V}_{\alpha}\tilde{V}_{\beta}
\xi_{\alpha}^{\nu}\xi_{\beta}^{\nu} \right]~~~.
\ee
We see that the  characteristic speed
\be
\bar{v}^{2}=(\Gamma c)^{2}\frac{\bar{S}_{4}}{\bar{S}_{2}}
\label{eq:97}
\ee
has been introduced into the problem.

We can construct a form  for the integral $I(\tilde{V})$ which does not
involve a unit vector in the $\delta$-function via the
following rearrangements:
\be
\delta (\hat{{\cal D}}\cdot\vec{\tilde{V}})=
|\vec{{\cal D}}|\delta (\vec{{\cal D}}\cdot\vec{\tilde{V}})
\nonumber
\ee
\be
=\frac{\vec{{\cal D}}^{2}}{|\vec{{\cal D}}|}
\delta (\vec{{\cal D}}\cdot\vec{\tilde{V}})
\nonumber
\ee
\be
=\frac{\vec{{\cal D}}^{2}}{\left( det N\right)^{1/2}}
\delta (\vec{{\cal D}}\cdot\vec{\tilde{V}})
\nonumber
\ee
\be
=\int ~\frac{d^{n}z}{\left(2\pi \right)^{n/2}} \vec{{\cal D}}^{2}
\delta (\vec{{\cal D}}\cdot\vec{\tilde{V}})
e^{-\frac{1}{2}z_{\nu}N_{\nu ,\nu '}z_{\nu '}} ~~~.
\ee
Inserting this result into the integral $I[\tilde{V}]$ gives
\be
I(\tilde{V})=
\int  \prod_{\mu=1}^{d}\prod_{\nu=1}^{n}d\xi_{\mu}^{\nu}
\int ~\frac{d^{n}z}{\left(2\pi \right)^{n/2}}
\vec{{\cal D}}^{4}(\xi )\delta (\vec{{\cal D}}(\xi )\cdot\vec{\tilde{V}})
\frac{e^{-\frac{1}{2}A_{0}(\xi )}}{(2\pi )^{nd/2}}
\label{eq:99}
\ee
where
\be
A_{0}(\xi , z )=
(\xi_{\alpha}^{\nu})^{2}
+\tilde{V}_{\alpha}\tilde{V}_{\beta}
\xi_{\alpha}^{\nu}\xi_{\beta}^{\nu}
+ z_{\nu}\xi_{\alpha}^{\nu}\xi_{\alpha}^ {\nu '}z_{\nu '}~~~.
\ee

\section{Case $\n =1$ and $\D =2$}

It is straightforward  to work out $P[{\bf V}]$ and $n_{0}$ for the
simplest case of defect lines in two dimensions.  The key simplifying
aspect in
this example is that the matrix $\xi_{\alpha}^{\nu}$ reduces to a
vector $\xi_{\alpha}$  and
\be
{\cal D}_{s_{1}}(\xi )=\epsilon_{s_{1}\mu_{1}}\xi_{\mu_{1}}
\ee
with
\be
\vec{{\cal D}}^{2}=\vec{\xi}^{2} ~~~.
\ee
We have from Eq.(\ref{eq:94}) that
\be
n_{0}P[{\bf V}]=
\frac{1}{\left(2\pi \right)}\frac{1}{(\Gamma c)^{2}}
\sqrt{\frac{\bar{S}_{2}}{\bar{S}_{4}}}
\left(\frac{\bar{S}_{2}^{2}}{S_{0} \bar{S}_{4}}\right)^{1/2}
I(\tilde{V})
\ee
\be
I(\tilde{V})=
\int  d^{2}\xi
\int ~\frac{dz}{\left(2\pi \right)^{1/2}}
\vec{\xi }^{4}\delta (\vec{{\cal D}}(\xi )\cdot\vec{\tilde{V}})
\frac{1}{(2\pi )}
e^{-\frac{1}{2}A_{0}(\xi )}
\ee
and
\be
A_{0}(\xi , z )=
\xi_{\alpha}^{2}
+\tilde{V}_{\alpha}\tilde{V}_{\beta}
\xi_{\alpha}\xi_{\beta}
+ z^{2}\xi_{\alpha}^{2}~~~.
\ee
Then, since the integral is isotropic, we can pick ${\bf V}$ to be
in the $x$-direction and use $\vec{{\cal D}}(\xi )\cdot\vec{\tilde{V}}
=V\xi_{y}$ in the $\delta$-function to do the integral over $\xi_{y}$
and obtain
\be
I(\tilde{V})=\frac{1}{(2\pi \tilde{V})}
\int  d\xi_{x}
\int ~\frac{dz}{\left(2\pi \right)^{1/2}}
\xi_{x} ^{4} ~
e^{-\frac{1}{2}\xi_{x}^{2}[1+z^{2}+\tilde{V}^{2}] } ~~~.
\ee
The remaining integrals are elementary and we obtain the results
\be
I(\tilde{V})=\frac{2}{(\pi \tilde{V})}
\frac{1}{(1+ \tilde{V}^{2})^{2}}
\ee
and
\be
n_{0}P[{\bf V}]=
\frac{1}{\left(2\pi \right)}\frac{1}{\Gamma c}
\sqrt{\frac{\bar{S}_{2}}{\bar{S}_{4}}}
\left(\frac{\bar{S}_{2}^{2}}{S_{0} \bar{S}_{4}}\right)^{1/2}
\frac{2}{(\pi \tilde{V})}
\frac{1}{(1+ \tilde{V}^{2})^{2}} ~~~.
\ee
We can then obtain the string density $n_{0}$ in two ways.  We can
compute it directly from
\be
n_{0}=<|\vec{{\cal D}}(\vec{\psi})|\delta (\vec{\psi})>
\ee
\be
=\int ~\prod_{\nu=1}^{n}d\xi_{\mu}^{\nu}
|\vec{{\cal D}} (\xi )|
G(\xi )
\ee
where $G(\xi )$ is the integral over $\vec{b}$ of Eq.(\ref{eq:62}) given by
\be
G(\xi )=\frac{1}{(2\pi S_{0})^{n/2}}
\frac{1}{(2\pi \bar{S}_{2})^{nd/2}}
exp\biggl[-\frac{1}{2\bar{S}^{(2)}}\sum_{\mu
,\nu}(\xi_{\mu}^{\nu})^{2}\biggr] ~~~.
\ee
Restricting the analysis to $n=1$ and $d=2$ gives
\be
n_{0}=\int ~d^{2}\xi \frac{1}{\sqrt{2\pi S_{0}}}
\frac{1}{2\pi \bar{S}_{4}}
e^{-\frac{1}{2\bar{S}_{2}}\vec{\xi}^{2}} ~~~.
\ee
These integrations are elementary with the final result:
\be
n_{0}=\frac{1}{2}\left(\frac{\bar{S}_{2}}{S_{0}}\right)^{1/2} ~~~.
\ee
The second method for determining $n_{0}$, which serves as a check on
intermediate steps in the calculation, is to integrate
$n_{0}P[{\bf V}]$ over all ${\bf V}$ and use the fact that
$P[{\bf V}]$ must be normalized.  The exercise is straightforward and
leads to the same result for $n_{0}$.  We then have the final result
for the probability distribution
\be
P[{\bf V}]=\frac{2}{\pi^{2}}\frac{1}{V\bar{v}}
\frac{1}{(1+(V/\bar{v})^{2})^{2}}
\ee
and it is easy to see that this agrees with Eq.(\ref{eq:1}) for
$n=1$ and $d=2$.

\section{Case $\n =2$ and $\D =3$}

We turn to the physically important case of strings in three spatial
dimensions.  In working out this case, the key observation is that
\be
{\cal D}_{\alpha} (\xi )=\frac{1}{2}\epsilon_{\alpha ,\mu_{1},\mu_{2}}
\epsilon_{\nu_{1},\nu_{2}}\xi_{\mu_{1}}^{\nu_{1}}\xi_{\mu_{2}}^{\nu_{2}}
{}~~~,
\ee
is a quadratic form in $\xi $.  This suggests that we use the
integral representation of the $\delta$-function
\be
\delta (\vec{{\cal D}}(\xi )\cdot\vec{\tilde{V}})=
\int ~\frac{dk}{2\pi}~e^{ik\vec{{\cal D}}(\xi )\cdot\vec{\tilde{V}}}
\ee
to write the integral of interest, Eq.(\ref{eq:99}), in the form:
\be
I(\tilde{V})=
\int  \prod_{\mu=1}^{3}\prod_{\nu=1}^{2}\frac{d\xi_{\mu}^{\nu}}
{(2\pi )^{3}}
\int ~\frac{d^{2}z}{2\pi } \int ~\frac{dk}{2\pi }
\vec{{\cal D}}^{4}(\xi )
e^{-\frac{1}{2}\xi_{\mu}^{\nu}Q_{\mu\mu'}^{\nu\nu'}\xi_{\mu'}^{\nu'}}
{}~~~.
\ee
The matrix appearing in the gaussian is given by
\be
Q_{\mu\mu'}^{\nu\nu'}
=\delta_{\mu\mu'}\delta_{\nu\nu'}
+\delta_{\nu\nu'}\tilde{V}_{\mu}\tilde{V}_{\mu'}
+\delta_{\mu\mu'}z_{\nu}z_{\nu'}
-ik_{\alpha}\epsilon_{\alpha ,\mu,\mu'}
\epsilon_{\nu,\nu'} ~~~,
\label{eq:118}
\ee
where we have introduced $k_{\alpha}=k\tilde{V}_{\alpha} $ and have used
the result
\be
-\frac{1}{2}\xi_{\mu}^{\nu}\left(-ik_{\alpha}\epsilon_{\alpha ,\mu,\mu'}
\right)\xi_{\mu'}^{\nu'}=i\vec{k}\cdot\vec{{\cal D}}(\xi ) ~~~.
\ee
We see that in principle we can carry out the $\xi $ integration since
it involves a product of polynomials times a gaussian weight.  Let us
define the integral
\be
L(\vec{k},z)=\int  \prod_{\mu=1}^{3}\prod_{\nu=1}^{2}\frac{d\xi_{\mu}^{\nu}}
{(2\pi )^{3}}\vec{{\cal D}}^{4}(\xi )
e^{-\frac{1}{2}\xi_{\mu}^{\nu}Q_{\mu\mu'}^{\nu\nu'}\xi_{\mu'}^{\nu'}}
\ee
and
\be
I(\tilde{V})=\int ~\frac{d^{2}z}{2\pi } \int ~\frac{dk}{2\pi }L(\vec{k},z)
{}~~~.
\ee
A significant simplification occurs if we realize that gradients with
respect to $\vec{k}$ pull down factors of $\vec{{\cal D}}$ and we can
write
\be
L(\vec{k},z)=\nabla_{k}^{4}L_{0}(\vec{k},z)
\ee
where
\be
L_{0}(\vec{k},z)=
\int  \prod_{\mu=1}^{3}\prod_{\nu=1}^{2}\frac{d\xi_{\mu}^{\nu}}
{(2\pi )^{3}}
e^{-\frac{1}{2}\xi_{\mu}^{\nu}Q_{\mu\mu'}^{\nu\nu'}\xi_{\mu'}^{\nu'}}
{}~~~.
\ee
Thus we are left with a set of gaussian integrals.

The gaussian integral can be carried out if we think of
$Q_{\mu\mu'}^{\nu\nu'}$ as a $6\times 6$ symmetric matrix.
If we can find the eigenvalues $\lambda_{i}, i=1,2\ldots,6$ of this matrix,
then we have
\be
L_{0}(\vec{k},z)=\frac{1}{\sqrt{\prod_{i=1}^{6}\lambda_{i}}} ~~~.
\ee
In Appendix E we discuss the relevant eigenvalue problem.  The result
of the analysis there is that the six-dimensional problem factorizes into a
product of two three-dimensional problems.  These three-dimensional  eigenvalue
problems reduce to cubic equations and the product of the three associated
eigenvalues can be read off from the associated characteristic equation
with the final result:
\be
L_{0}(\vec{k},z)=\frac{1}{\sqrt{(1+z^{2})(1+z^{2}+\tilde{V}^{2}+k^{2})
+(\vec{k}\cdot\vec{\tilde{V}})^{2}}}
\frac{1}{\sqrt{(1+z^{2})(1+\tilde{V}^{2})+k^{2}
+(\vec{k}\cdot\vec{\tilde{V}})^{2}}} ~~~.
\ee
We must then apply $\nabla_{k}^{4}$ to $L_{0}(\vec{k},z)$
and set $\vec{k}=k\vec{\tilde{V}}$.
After a great deal of algebra one finds a complicated result for
$I[\tilde{V}]$ which still requires integration over $k$ and $\vec{z}$.
It turns out that it is wise to first do the
$k$ integration.  All of the
contributions are proportional to integrals of the form
\be
\int ~\frac{dk}{2\pi}\frac{(k\tilde{V})^{2p}}
{(1+z^{2}+\tilde{V}^{2}k^{2})^{p+3}}=
\frac{\kappa_{p}}{2\pi \tilde{V}}
\frac{1}{(1+z^{2})^{5/2}}
\ee
where $p$ takes the values $0,1$ and $2$ with
$\kappa_{0}=\frac{3\pi}{8}$, $\kappa_{1}=\frac{\pi}{16}$,
$\kappa_{0}=\frac{3\pi}{8(16)}$.   The final integrals over $\vec{z}$ can
all be expressed in terms of the integrals
\be
J_{s_{1},s_{2}}=\int ~ \frac{d^{2}z}{2\pi}
\frac{1}{(1+z^{2}+\tilde{V}^{2})^{s_{1}/2}}
\frac{1}{(1+z^{2})^{s_{2}/2}}
\ee
which, for integer $s_{1}$ and $s_{2}$, can be worked out analytically.
After an enormous amount of additional algebra we obtain the very simple result
\be
I[\tilde{V}]=\frac{3}{\tilde{V}}
\frac{1}{(1+\tilde{V}^{2})^{5/2}} ~~~.
\ee
This leads back to the result
\be
n_{0}P[{\bf V}]=\frac{1}{(2\pi )^{2}}\frac{1}{(\Gamma c)^{3}}
\sqrt{\frac{\bar{S}_{2}}
{\bar{S}_{4}}} \frac{\bar{S}_{2}^{2}}{S_{0}\bar{S}_{4}}
\frac{3}{\tilde{V}}
\frac{1}{(1+\tilde{V}^{2})^{5/2}} ~~~.
\label{eq:129}
\ee
Integrating over all ${\bf V}$ we easily obtain the density of strings
\be
n_{0}=\frac{1}{\pi}\frac{\bar{S}_{2}}{S_{0}} ~~~.
\ee
Putting this result back into Eq.(\ref{eq:129}) gives the final result
\be
P[{\bf V}]=\frac{3}{4\pi}\frac{1}{\bar{v}^{3}\tilde{V}}
\frac{1}{(1+\tilde{V}^{2})^{5/2}}
\ee
which agrees with Eq.(\ref{eq:1}) for $n=2$ and $d=3$.

\section{Walls in Three Dimensions}

Let us conclude with the example of walls in three dimensions
($n=1,d=3$).  We have from Section V that
\be
{\cal D}^{2}=\left(\vec{\nabla}\psi \right)^{2}
\ee
and the velocity of the wall is given by
\be
v_{\mu}=-\frac{\nabla_{\mu}\psi }{\left(\vec{\nabla}\psi \right)^{2}}
\Gamma c \nabla^{2}\psi ~~~.
\ee
Following the same path as for the point and string
defects we find that the velocity probability distribution is given by
\be
n_{0}P[{\bf V}]=~\int db ~d^{3}\xi|\vec{\xi}|
\delta \left(\vec{V}+\Gamma c\vec{\xi}b/\xi^{2}\right) G(\vec{\xi},b)
\ee
where $G(\vec{\xi},b)$ is given  by Eq.(\ref{eq:62}) with $n=1$ and
$d=3$ and

\be
n_{0}P[{\bf V}]=~\int db ~d^{3}\xi|\vec{\xi}|
\delta \left(\vec{V}+\Gamma c\vec{\xi}b/\xi^{2}\right)
\frac{1}{(2\pi S_{0})^{1/2}}
\frac{e^{-\frac{1}{2\bar{S}_{4}}\vec{b}^{2}}}{(2\pi \bar{S}_{4})^{1/2}}
\frac{e^{-\frac{1}{2\bar{S}_{2}}\vec{\xi}^{2}}}{(2\pi \bar{S}_{2})^{d/2}}
\ee
Again inserting the integral representation for the $\delta$-function
and doing the $b$ integration, we obtain
\be
n_{0}P[{\bf V}]= \frac{1}{\sqrt{2\pi S_{0}(2\pi \bar{S}_{2})^{d}}}
\int ~d^{3}\xi|\vec{\xi}|
exp\biggl[-\frac{1}{2\bar{S}_{2}}\vec{\xi}^{2}\biggr] J(\xi )
\ee
where
\be
J(\xi )=\int \frac{d^{3}k}{(2\pi )^{3}}
e^{i\vec{k}\cdot\vec{V}}
exp\left[-\frac{1}{2}\bar{S}_{4}(\Gamma c)^{2}(\vec{k}\cdot\hat{\xi})^{2}
/\xi^{2}\right] ~~~.
\label{eq:132}
\ee
The integral in Eq.(\ref{eq:132}) can be carried out if we introduce the
orthogonal set of coordinates $\hat{\xi}$, $\hat{b}^{(1)}$,
$\hat{b}^{(2)}$ and we choose
\be
\hat{\xi}=\hat{b}^{(1)}\times \hat{b}^{(2)} ~~~.
\ee
Then we obtain
\be
J(\xi )=\delta (\vec{V}\cdot\hat{b}^{(1)})
\delta (\vec{V}\cdot\hat{b}^{(2)})
\frac{|\vec{\xi}|}{\sqrt{2\pi \bar{S}_{4}(\Gamma c)^{2}}}
exp\left[-\frac{1}{2\bar{S}_{4}(\Gamma c)^{2}}
(\vec{V}\cdot\vec{\xi})^{2}\right] ~~~.
\ee
If we make the same change of variables given by
Eqs.(\ref{eq:91}) and (\ref{eq:92})
we find the result
\be
n_{0}P[{\bf V}]=\frac{1}{\bar{v}^{3}}\frac{1}{(2\pi )^{5/2}}
\sqrt{\frac{\bar{S}_{2}}{\bar{S}_{4}}}I[\tilde{V}]
\label{eq:140}
\ee
and
\be
I[\tilde{V}]=
\int ~d^{3}\xi ~\vec{\xi}^{2}
\delta (\tilde{\vec{V}}\cdot\hat{b}^{(1)})
\delta (\tilde{\vec{V}}\cdot\hat{b}^{(2)})
e^{-\frac{1}{2}\left[\vec{\xi}^{2}
+(\vec{\xi}\cdot\tilde{\vec{V}})^{2}\right]}
{}~~~.
\ee
If we now define
\be
\vec{b}^{(1)}=\hat{b}^{(2)}\times\vec{\xi}
\ee
and
\be
\vec{b}^{(2)}=\vec{\xi}\times\hat{b}^{(2)}
\ee
and assume that $\vec{V}$ is in the $z$-direction we obtain
\be
I[\tilde{V}]=
\int ~d^{3}\xi\vec{\xi}^{4}
\delta (\tilde{V}b^{(1)}_{z})
\delta (\tilde{V}b^{(2)}_{z})
e^{-\frac{1}{2}\left[\vec{\xi}^{2}
+\xi_{z}^{2}\tilde{V}^{2}\right]}
{}~~~.
\ee
The key point then is that
\be
\delta (\tilde{V}b^{(1)}_{z})
\delta (\tilde{V}b^{(2)}_{z})=\frac{1}{\tilde{V}^{2}}
\delta (\xi_{x})\delta (\xi_{y})
\ee
and the integral reduces to an elementary  one-dimensional integral
\be
I[\tilde{V}]=\frac{1}{\tilde{V}^{2}}
\int ~d\xi_{z}\xi_{z}^{4}
e^{-\frac{1}{2}\vec{\xi}^{2}\left[
1+\tilde{V}^{2}\right]}
{}~~~.
\ee
Inserting the result of doing this integral back into Eq.(\ref{eq:140})
leads
immediately to the result that
$P[{\bf V}]$ is again given by Eq.(\ref{eq:1})~~.

\section{Determination of $\bar{\V }$}

The determination of $S_{0},\bar{ S}_{2}$,
$\bar{S}_{4}$ and $\bar{v}$ requires a theory for the auxiliary field
correlation
function
\be
C_{0}(12)=\frac{1}{n}\langle \vec{m}(1)\cdot \vec{m}(2)\rangle ~~~~.
\ee
If we follow the development in Refs.[\onlinecite{EXP}]and [\onlinecite{ctg}] we can write in
the scaling regime,
\be
 C_{0}(12)=S_{0}e^{-\vec{r}^{2}/(2L^{2})}
\ee
where $\vec{r}=\vec{r_{1}}-\vec{r}_{2}$.  We do not need
$S_{0},\bar{ S}_{2}$, and
$\bar{S}_{4}$ individually.  We only need the combination given by
Eq.(\ref{eq:97}).  We easily find
\be
\bar{v}^{2}=\frac{2d}{L^{2}}(\Gamma c )^{2}=\frac{\Gamma c d}{2t} ~~~.
\ee

\section{Conclusions and Questions}

The results of this work are simply summarized by Eq.(\ref{eq:1}).
There remain several open questions.  The most important question is
whether Eq.(\ref{eq:1}) for $P[{\bf V}]$ corresponds to the results found in
the real world.  It is highly desirable to measure $P[{\bf V}]$
for the rather broad range of systems covered by Eq.(\ref{eq:1}).

How is  this result for the defect velocity probability distribution
changed at higher order
in perturbation theory?
So far,
$P({\bf V})$ has only
been calculated for the lowest order in
the perturbation theory developed in Ref.(\onlinecite{EXP}).
Scaling arguments\cite{bvvt} would indicate that the large velocity tails
will not be modified at
higher orders in perturbation theory.  This question will be addressed
soon.

The results for $P[{\bf V}]$ given by Eq.(\ref{eq:1}) have only been
proven for a subset of the range $n\le d$.  What about for
$n=d-2=2$?  This question is somewhat academic since it is outside the
range of
physically accessible spatial dimensions.

%
%
\acknowledgments

This work was supported primarily by the MRSEC Program of the National Science
Foundation under Award Number DMR-9400379.

\appendix

\section{$\vec{{\cal D}}^{2}$}

Clearly one of the important quantities entering into the discussion of
the string velocity probability distribution is the quantity
${\cal D}$ defined by Eq.(\ref{eq:20}) and its square:
\be
\vec{{\cal D}}^{2}=
\frac{1}{(n!)^{2}}
\epsilon_{\mu\mu_{1}\mu_{2}\ldots\mu_{n}}
\epsilon_{\nu_{1}\nu_{2}\ldots\nu_{n}}
\xi_{\mu_{1}}^{\nu_{1}}
\xi_{\mu_{2}}^{\nu_{2}}\ldots
\xi_{\mu_{n}}^{\nu_{n}}
\epsilon_{\mu ,\mu_{1}'\mu_{2}'\ldots\mu_{n}'}
\epsilon_{\nu_{1}'\nu_{2}'\ldots\nu_{n}'}
\xi_{\mu_{1}'}^{\nu_{1}'}
\xi_{\mu_{2}'}^{\nu_{2}'}\ldots
\xi_{\mu_{n}'}^{\nu_{n}'} ~~~.
\ee
This quantity cn be written in a simplified form if we realize that
\be
\epsilon_{\mu\mu_{1}\mu_{2}\ldots\mu_{n}}
\epsilon_{\mu\mu_{1}'\mu_{2}'\ldots\mu_{n}'}
=I[\mu_{1}\mu_{2}\ldots\mu_{n};\mu_{1}'\mu_{2}'\ldots\mu_{n}']
\label{eq:A1}
\ee
where $I$  is a set of products of $\delta$-functions giving all matched
pairs between the unprimed and primed sets.  There are minus signs
if the matched pairs are an odd number of permutations of the labels
in order to return to the order $1,2,\ldots,n$.   We have then
\be
\vec{{\cal D}}^{2}=
\frac{1}{(n!)^{2}}
I[\mu_{1}\mu_{2}\ldots\mu_{n};\mu_{1}'\mu_{2}'\ldots\mu_{n}']
\epsilon_{\nu_{1}\nu_{2}\ldots\nu_{n}}
\epsilon_{\nu_{1}'\nu_{2}'\ldots\nu_{n}'}
\xi_{\mu_{1}}^{\nu_{1}}
\xi_{\mu_{2}}^{\nu_{2}}\ldots
\xi_{\mu_{n}}^{\nu_{n}}
\xi_{\mu_{1}'}^{\nu_{1}'}
\xi_{\mu_{2}'}^{\nu_{2}'}\ldots
\xi_{\mu_{n}'}^{\nu_{n}'}
\nonumber
\ee
\be
=\frac{1}{n!}
\epsilon_{\nu_{1}\nu_{2}\ldots\nu_{n}}
\epsilon_{\nu_{1}'\nu_{2}'\ldots\nu_{n}'}
N_{\nu_{1}\nu_{1}'}N_{\nu_{2}\nu_{2}'}\ldots
N_{\nu_{n}\nu_{n}'}
\nonumber
\ee
\be
=\frac{1}{n!}
\epsilon_{\nu_{1}\nu_{2}\ldots\nu_{n}}
\epsilon_{\nu_{1}\nu_{2}\ldots\nu_{n}}det (N)
\nonumber
\ee
\be
=det (N) ~~~.
\label{eq:A3}
\ee

\section{Matrix $M_{\mu}^{\nu}$}

The matrix $M_{\mu}^{\nu}$ defined by Eq.(\ref{eq:67}) can be put into
another useful form by inserting the explicit form for $\vec{{\cal D}}$
given by Eq.(\ref{eq:60}) and using the result given by Eq.(\ref{eq:A1}).  An
intermediate step gives
\be
\epsilon_{\mu \mu_{1}\mu_{2}\ldots\mu_{n}}{\cal D}_{\mu_{1}}
=\epsilon_{\mu \mu_{1}\mu_{2}\ldots\mu_{n}}
\frac{1}{n!}
\epsilon_{\mu_{1}\mu_{1}'\mu_{2}'\ldots\mu_{n}'}
\epsilon_{\nu_{1}'\nu_{2}'\ldots\nu_{n}'}
\xi_{\mu_{1}'}^{\nu_{1}'}
\xi_{\mu_{2}'}^{\nu_{2}'}\dots
\xi_{\mu_{n}'}^{\nu_{n}'}
\nonumber
\ee
\be
=-I[\mu,\mu_{2}\ldots\mu_{n};\mu_{1}'\mu_{2}'\ldots\mu_{n}']
\frac{1}{n!}\epsilon_{\nu_{1}'\nu_{2}'\ldots\nu_{n}'}
\xi_{\mu_{1}'}^{\nu_{1}'}
\xi_{\mu_{2}'}^{\nu_{2}'}\ldots
\xi_{\mu_{n}'}^{\nu_{n}'}
\nonumber
\ee
\be
=-\epsilon_{\nu_{1}'\nu_{2}'\ldots\nu_{n}'}
\xi_{\mu}^{\nu_{1}'}
\xi_{\mu_{2}}^{\nu_{2}'}\ldots
\xi_{\mu_{n}}^{\nu_{n}'} ~~~.
\ee
Inserting this into the definition of $M_{\mu}^{\nu}$ gives
\be
M_{\mu}^{\nu}=
-\frac{1}{\vec{{\cal D}}^{2}}\frac{1}{(n-1)!}
\epsilon_{\nu\nu_{2}\ldots\nu_{n}}
\xi_{\mu_{2}}^{\nu_{2}}\ldots
\xi_{\mu_{n}}^{\nu_{n}}
\epsilon_{\nu_{1}'\nu_{2}'\ldots\nu_{n}'}
\xi_{\mu_{1}'}^{\nu_{1}'}
\xi_{\mu_{2}'}^{\nu_{2}'}\ldots
\xi_{\mu_{n}'}^{\nu_{n}'}
\nonumber
\ee
\be
=-\frac{1}{\vec{{\cal D}}^{2}}\frac{1}{(n-1)!}
\epsilon_{\nu\nu_{2}\ldots\nu_{n}}\xi_{\mu}^{\nu_{1}'}
\epsilon_{\nu_{1}'\nu_{2}'\ldots\nu_{n}'}
N_{\nu_{2}\nu_{2}'}N_{\nu_{3}\nu_{3}'}\ldots N_{\nu_{n}\nu_{n}'}
\ee
This  expression is particularly useful if we consider the quantity
\be
\xi_{\mu}^{\nu'}M_{\mu}^{\nu}
=-\frac{1}{\vec{{\cal D}}^{2}}\frac{1}{(n-1)!}
\epsilon_{\nu,\nu_{2}\ldots\nu_{n}}
\epsilon_{\nu_{1}'\nu_{2}'\ldots\nu_{n}'}
N_{\nu'\nu_{1}'}N_{\nu_{2}\nu_{2}'}\dots
N_{\nu_{n}\nu_{n}'}
\nonumber
\ee
\be
=-\frac{1}{\vec{{\cal D}}^{2}}\frac{1}{(n-1)!}
\epsilon_{\nu,\nu_{2}\ldots\nu_{n}}
\epsilon_{\nu',\nu_{2}\ldots\nu_{n}} det ~ N
\nonumber
\ee
\be
=-\delta_{\nu ,\nu'} ~~~.
\label{eq:121}
\ee

\section{Orthogonal Coordinate System}

We will construct an orthonormal coordinate system with the basis
vectors
\be
\hat{\xi}_{\alpha}^{(s)}=
\sum_{\nu =1}^{n} A_{s\nu}\xi _{\alpha}^{\nu}  ~~~.
\label{eq:122}
\ee
These basis vectors are orthogonal to $\vec{{\cal D}}$ since
\be
\xi _{\alpha}^{\nu}{\cal D}_{\alpha}=0 ~~~.
\ee
Since we require
\be
\hat{\xi}_{\alpha}^{(s)}\hat{\xi}_{\alpha}^{(s')}=\delta_{ss'} ~~~,
\label{eq:124}
\ee
we have immediately, on inserting Eq.(\ref{eq:122}) into
Eq.(\ref{eq:124}), that
\be
\delta_{ss'}=A_{s\nu}A_{s'\nu'}N_{\nu\nu'} ~~~.
\ee
If we take the determinant we obtain
\be
1= det ~ A ~det ~ N ~det ~\tilde{A}
\ee
or, using Eq.(\ref{eq:A3}),
\be
det ~\tilde{A}~  det ~ A =\frac{1}{det ~ N}=\frac{1}{\vec{{\cal D}}^{2}}
{}~~~.
\label{eq:C6}
\ee

We can obtain  a realization of the matrix
$A_{s\nu}$  by constructing the $\hat{\xi}_{\alpha}^{(s)}$
directly
using the basis set
\be
\chi_{\alpha}^{(s)}=\sum_{\nu =1}^{n}
e^{2\pi i s\nu/n} \xi _{\alpha}^{\nu}
\ee
which are not orthonormal.  However they can be constructed to be
orthonormal using the Gram-Schmidt orthogonalization process.
Thus each $\hat{\xi}_{\alpha}^{(s)}$ is a linear combination of the
$\chi_{\alpha}^{(s)}$ which is proportional to a linear combination
of $\xi _{\alpha}^{\nu}$.  Thus one can extract an explicit expression
for the matrix $A$.  This explicit expression is not needed here.

\section{Inverse of Matrix Q}

We need the inverse of the matrix
\be
Q_{\nu\nu'}=\hat{\xi}_{\alpha}^{(\nu )}\hat{\xi}_{\beta}^{(\nu' )}
\bar{M}_{\alpha\beta}~~~.
\ee
If we use the expression given by Eq.(\ref{eq:122}) for the basis vectors
we obtain
\be
Q_{\nu\nu'}=
A_{\nu\bar{\nu}}\xi_{\alpha}^{\bar{\nu}}
A_{\nu'\bar{\nu}'}\xi_{\beta}^{\bar{\nu}'}
M_{\alpha}^{\nu''}M_{\beta}^{\nu''} ~~~.
\ee
We then use the identity Eq.(\ref{eq:121}) twice to obtain
\be
Q_{\nu\nu'}=A_{\nu\bar{\nu}} A_{\nu'\bar{\nu}'}
\delta_{\nu'',\bar{\nu}}\delta_{\nu'',\bar{\nu}'}
=A_{\nu\bar{\nu}} A_{\nu'\bar{\nu}} ~~~.
\ee
This immediately tells us that
\be
det ~Q =det ~A ~ det ~\tilde{A}=\frac{1}{\vec{{\cal D}}^{2}} ~~~.
\ee
We still need to construct the inverse of $Q$.  It is easy to see that
this is given by the product
\be
Q^{-1}=\tilde{A}^{-1} A^{-1}  ~~~.
\ee
However we only need the elements
\be
R_{\alpha\beta}=\hat{\xi}_{\alpha}^{(\nu )}\left(Q^{-1}\right)_{\nu\nu'}
\hat{\xi}_{\beta}^{(\nu' )}
\nonumber
\ee
\be
=A_{\nu s}\xi_{\alpha}^{s}\left(\tilde{A}^{-1}\right)_{\nu\bar{\nu}}
\left(A^{-1}\right)_{\bar{\nu}\nu '}A_{\nu' s'}\xi_{\beta}^{s'}
\nonumber
\ee
\be
=\xi_{\alpha}^{s}\delta_{\bar{\nu},s}\delta_{\bar{\nu},s'}\xi_{\beta}^{s'}
\nonumber
\ee
\be
=\xi_{\alpha}^{s}\xi_{\beta}^{s} ~~~.
\ee

\section{Eigenvalue Problem}

We need to find the six eigenvalues for the matrix $Q$ defined by
Eq.(\ref{eq:118}).  The analysis can be carried out by looking at the action
of $Q$ when acting on the six basis vectors:
$\psi_{1}=\hat{V}_{\mu}\hat{z}_{\nu}, \psi_{2}=\hat{V}_{\mu}\hat{c}_{\nu},
\psi_{3}=\hat{b}_{\mu}\hat{z}_{\nu}, \psi_{4}=\hat{b}_{\mu}\hat{c}_{\nu},
\psi_{5}=\hat{k}_{\mu}\hat{z}_{\nu}, \psi_{6}=\hat{k}_{\mu}\hat{c}_{\nu}$
where we have introduced $\hat{c}_{\nu}=\epsilon_{\nu\nu'}\hat{z}_{\nu'}$,
and $\hat{b}=\hat{k}\times\hat{V}$.  The action of $Q$ acting on these
states is given by
\be
Q\psi_{1}=(1+z^{2}+\tilde{V}^{2})\psi_{1}
+ik\psi_{4}
\ee
\be
Q\psi_{2}=(1+\tilde{V}^{2})\psi_{2}
-ik\psi_{3}
\ee
\be
Q\psi_{3}=(1+z^{2})\psi_{3}
-ik\psi_{2} +i\vec{k}\cdot\hat{V}\psi_{6}
\ee
\be
Q\psi_{4}=\psi_{4}
+ik\psi_{1} -i\vec{k}\cdot\hat{V}\psi_{5}
\ee
\be
Q\psi_{5}=(1+z^{2})\psi_{5}
+\tilde{V}^{2}\hat{k}\cdot\hat{V}\psi_{1}
\ee
\be
Q\psi_{6}=\psi_{6}
+\tilde{V}^{2}\hat{k}\cdot\hat{V}\psi_{2} ~~~.
\ee
We see that these equations decouple into two cubic systems
$(\psi_{1},\psi_{4},\psi_{5})$, and $(\psi_{2},\psi_{3},\psi_{6})$.
We then only need the products $\lambda_{1}\lambda_{2}\lambda_{3}$ and
$\lambda_{4},\lambda_{5}\lambda_{6}$.  It is well known
(see Ref.[\onlinecite{HB}]) that if one has a cubic characteristic equation
\be
\lambda^{3}+a_{2}\lambda^{2}+a_{1}\lambda +a_{0}=0 ~~~,
\ee
then the product of the three roots is given by
\be
\lambda_{1}\lambda_{2}\lambda_{3}=-a_{0} ~~~.
\ee
In the present case we easily find for the two sets:
\be
a_{0}^{(1)}=-(1+z^{2})(1+z^{2}\tilde{V}^{2}+k^{2})
-(\vec{k}\cdot\vec{\tilde{V}})^{2}
\ee
and
\be
a_{0}^{(2)}=-(1+z^{2})(1+\tilde{V}^{2})1s-+k^{2})
-(\vec{k}\cdot\vec{\tilde{V}})^{2}  ~~~.
\ee

\pagebreak

\end{document}